\documentclass[namedreferences]{kluwer}
\usepackage{epsfig}

\begin{document}
\begin{article}

\begin{opening}

\title{Modeling Dust on Galactic SED:\\
Application to Semi--Analytical Galaxy Formation Models}

\author{Laura \surname{Silva} \email{silva@sissa.it}}
\institute{International School for Advanced Studies, Trieste, Italy }
\author{Gian Luigi \surname{Granato} \email{granato@pd.astro.it}}
\author{Alessandro \surname{Bressan} \email{bressan@pd.astro.it}}
\institute{Osservatorio Astronomico di Padova, Padova, Italy }
\author{Cedric \surname{Lacey} \email{lacey@tac.dk}}
\institute{Theoretical Astrophysics Center, Copenhagen, Denmark}
\author{Carlton M. \surname{Baugh} \email{c.m.baugh@durham.ac.uk}}
\author{Shaun \surname{Cole} \email{shaun.cole@durham.ac.uk}}
\author{Carlos S. \surname{Frenk} \email{c.s.frenk@durham.ac.uk}}
\institute{Physics Department, University of Durham, Durham, UK}

\begin{abstract}
We present the basic features and preliminary results of the interface between
our spectro--photometric model GRASIL (that calculates galactic SED from the UV
to the sub--mm with a detailed computation of dust extinction and thermal
reemission) with the semi--analytical galaxy formation model GALFORM (that
computes galaxy formation and evolution in the hierarchical scenario, providing
the star formation history as an input to our model). With these two models we
are able to synthetize simulated samples of a few thousands galaxies suited for
statistical studies of galaxy properties to investigate on galaxy formation and
evolution. We find good agreement with the available data of SED and
luminosity functions.
\end{abstract}

\end{opening}

\section{Introduction}

Dust in galaxies extinguishes starlight and thermally radiates it in the
FIR--submm. Observations indicate that a significant fraction of the high--z
star formation is hidden in the optical by dust reprocessing as observed in
local starbursts. Thus dust shapes the spectral energy distribution (SED) of
star forming galaxies as a function of its optical properties and of the
geometry of the star--dust distribution, i.e.\ to infer information on the
evolutionary status of galaxies from their spectra, a complex of very uncertain
radiative processes connected to dust must be taken into account. A simplified
treatment can lead to wrong estimates of fundamental quantities such as the star
formation rate (SFR).

To investigate on galaxy formation and evolution we have developed a
spectro--photometric model (GRASIL, see Silva et al.\ 1998 for details and
Granato et al., these proceedings) properly computing the radiative effects of
dust on starlight, in order to interpret the huge amount of available and
incoming data over a 4 decades spectral range.

We present here the application of our model as the observational interface to
the Durham group semi--analytical hierarchical galaxy formation model (GALFORM,
see Cole et al.\ 1994, Baugh et al.\ 1998 for details).
Basically, this model calculates the formation and evolution of dark matter halos
starting from an assumed cosmology and initial spectrum of density fluctuations,
then the process of galaxy formation within these evolving halos is calculated
using a set of simple, physically--motivated rules to model gas cooling, star
formation and supernova feedback. The output from GALFORM, star formation
histories and most geometrical parameters, are used as input by GRASIL to
calculate detailed SED for simulated samples of a few thousands galaxies at
different redshifts.

In the following sections we outline a brief description of the interface
between the two models and some preliminary results of the comparisons with
observations. The detailed description will be presented in forthcoming 
papers (Granato et al.\ 1999, Lacey et al.\ 1999).

\section{Model interface}

Our photometric model GRASIL has been implemented in several ways in order to
provide a proper observational interface to the semi--analitycal models. Two
aspects required particular consideration:

\begin{enumerate}[]
\item

In the hierarchical clustering scenario galaxies are built up from sub--units
which underwent independent star formation and hence chemical histories before
merging to form a galaxy at a given cosmic epoch. Therefore stars in the final
object do not have a metallicity univocally determined by their age, as commonly
assumed in monolithic one--zone chemical evolution models, but rather a broad
distribution for each age (see for instance Fig.~\ref{plot3d}, where the star
formation history of a sample model is shown as a function of time and
metallicity).

\begin{figure}
\centerline{\epsfig{file=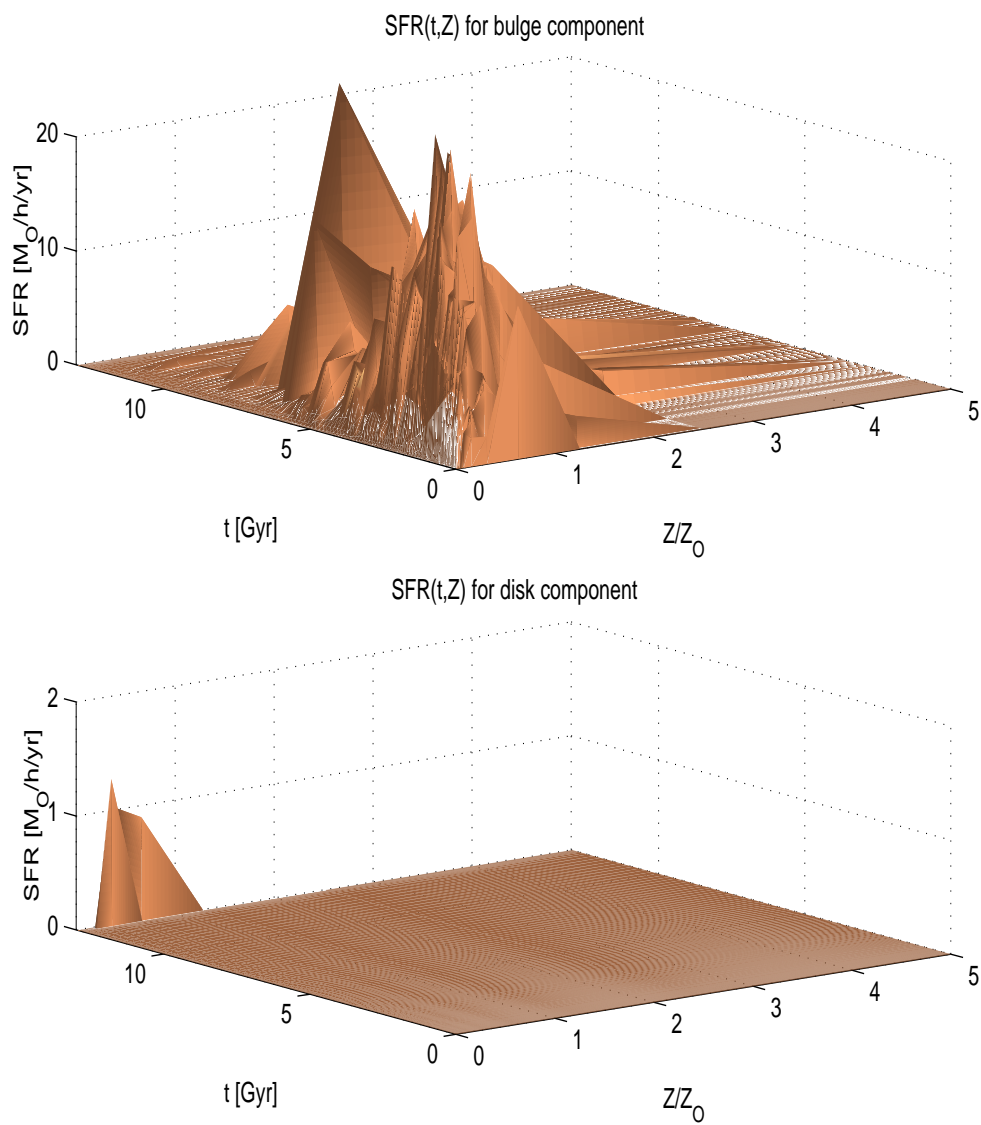,height=15truecm,width=13truecm}} 
\caption{An example of the star formation history calculated by the semi--analytical models:
the evolution of the different sub--units merging into a galaxy is represented
as a metallicity binned SFR, separately for the components merging into the bulge
and disk of the final galaxy.}
\label{plot3d}
\end{figure}

In order to compute the spectra of these composite objects, the semianalytical
code provides the evolution of the star--formation rate, $SFR_{z_j}(t_i)$,
subdivided into 11 logarithmic metallicity bins $z_j$ from log$z_1=-4$ to
log$z_{11}=-1$ in steps of $0.3$.

Hence we compute the average spectrum of the stellar population born at any
galactic time step $t_i$, $SP_\lambda(t_i)$, through a weighted mean of the
spectra of the single stellar populations (SSP) of age $t_{gal}-t_i$ and
metallicity $z_j, j=1,11$, with weights provided by the metallicity binned
SFR, as follows:

\begin{equation}
SP_\lambda(t_i)=
\frac{\sum_j SFR_{z_j}(t_i) \times SSP_\lambda(t_{gal}-t_i,z_j)}{\sum_j SFR_{z_j}(t_i)}
\end{equation}

\item

In Silva et al.\ (1998) we considered systems which are either spheroidal or
disk dominated. Now we have to properly describe galaxies of all morphological
classes, covering the range of bulge--to--disk ratios $B/D$ yielded by the
semi-analytical models in order to reproduce the distribution of morphologies.
The semi--analytical model provides the star--formation histories separately for
each component (see Fig.~\ref{plot3d}), we implemented a general composite
geometry in which stars and gas are distributed in both a disk--like component
and in a spheroidal one. 

To properly compute dust extinction and emission in this geometry in a
reasonable CPU time we optimized the angular and radial grids that define the
galaxy volume elements. Since gas is always in the disk component we adopted a
$\theta$ grid suited for disks, finer near the equatorial plane, as a
function of the disk flattening, i.e.\ of the ratio of the disk scale--height to
the galaxy radius, $z_{d}/r_{gal}$. The best radial grid was obtained by merging
one suited for bulges, dominating the inner part, with one for disks. Both were
obtained by imposing $\rho(r_i)/\rho(r_{i+1})=$const, with $\rho(r)$ given by a
King and an exponential profile respectively. This yields a radial array
dependent on {\it scale--radius}$/r_{gal}$, with {\it scale--radius} equal to
core and disk scale--radius.

\end{enumerate}

In summary, with these improvements our model is now {\it multicomponent} as far
as both the stellar mix as a function of time and the geometry are concerned.

\section{Preliminary results}

\begin{figure}[H]
\centerline{\epsfig{file=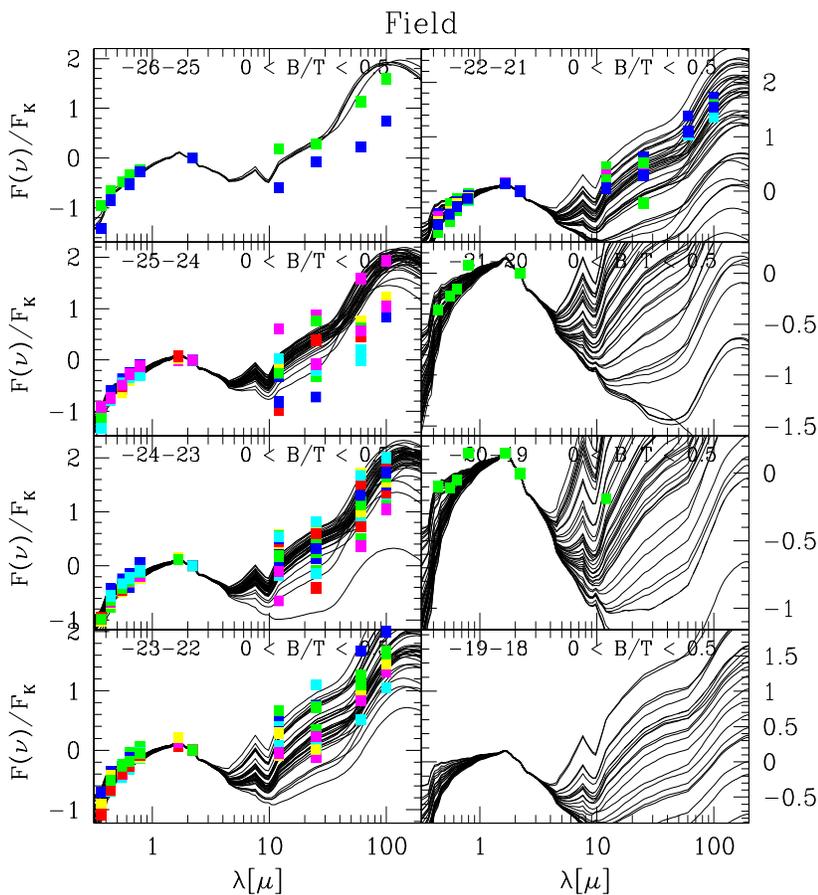,height=13cm,width=12cm}}
\caption{A sample of model SED for field face--on galaxies, binned in luminosity and B/T,
with $0<$ B/T $<0.5$,
of the galaxy catalogue generated by the semi--analytical model. They are
compared to available data for face--on galaxies in the same bins.}\label{figfk}
\end{figure}

A first check of the model is provided by the direct comparison of the SED of
the model galaxy catalogue with observed spectra, as the example shown in
Fig.~\ref{figfk} for field face--on galaxies with $0<$ B/T $<0.5$. For the luminosity and
B/T bins where enough data are available the comparison is statistically good
and well reproduces the great dispersion of the SED in the IR.

\begin{figure}[H]
\vspace{-1.5cm}
\centerline{\epsfig{file=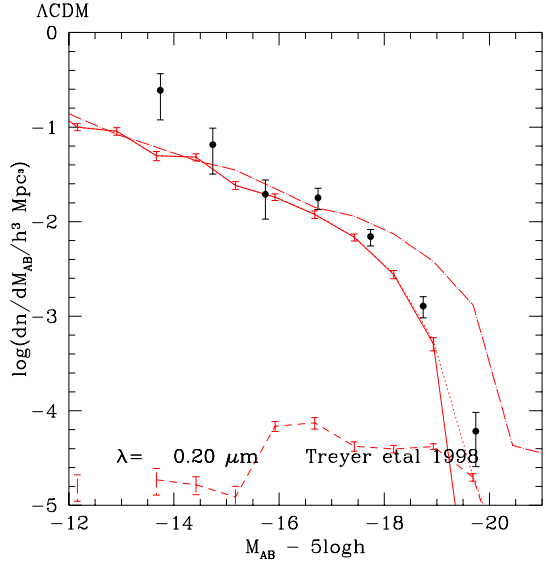,height=7.5cm,width=7.5cm}}
\caption{The $2000 \AA$ luminosity function. The solid line shows the model LF 
for galaxies without recent or ongoing bursts, while the dashed line shows the 
contribution of galaxies with recent or ongoing bursts. The dotted line shows
the total LF obtained by summing these two contributions. The dot--dashed line 
shows the total LF obtained if absorption by dust is neglected. Data are from 
a UV--selected redshift survey by Treyer et al.\ (1998).}\label{lf02}
\end{figure}

\begin{figure}[H]
\centerline{\epsfig{file=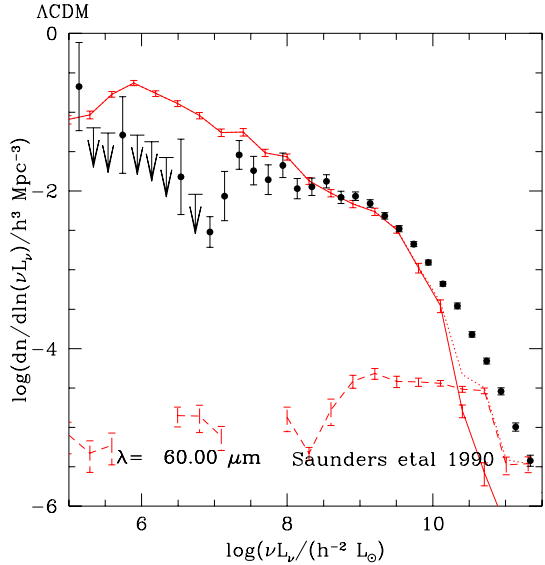,height=7.5cm,width=7.5cm}}
\caption{The $60 \mu$m luminosity function. The solid line shows the model LF
for galaxies without recent or ongoing bursts, while the dashed line shows the 
contribution of galaxies with recent or ongoing bursts. The dotted line shows
the total LF obtained by summing these two contributions. Data are from 
galaxies observed by IRAS by Saunders et al.\ (1990).}\label{lf60}
\end{figure}

The synthetic luminosity functions (LF) at different wavelenghts of the galaxy
catalogue have been computed and compared with observations. The LF at $0.2
\mu$m and $60 \mu$m are shown in Fig.~\ref{lf02} and \ref{lf60} respectively
(see captions for explanations). These are computed for certain prescriptions
for the cosmology and the star formation. In particular, the deficiency of the
$60 \mu$m LF at the high luminosity end is to be attributed mostly to the brown
dwarfs rich IMF adopted for these models rather than to an underestimate of the
UV--optical light dust absorption, since also the $0.2 \mu$m LF is slightly
scarce. Another possibility to be checked is that the number of galaxies with
ongoing or recent bursts is underestimated in the galaxy catalogue. We are
working to explore the range of parameters of GALFORM and GRASIL able to fit all
the available observational constraints.

\end{article}

\end{document}